 \documentclass[final,5p,times,twocolumn,authoryear]{elsarticle}

\usepackage{amssymb}
\usepackage{lipsum}
\journal{High Energy Astrophysics}
\usepackage{hyperref}
\hypersetup{
    colorlinks=true,
    linkcolor=myblue,
    citecolor=myblue,
    urlcolor=myblue
}
\usepackage{graphicx}
\usepackage{amsmath}
\usepackage[pagewise]{lineno}
\usepackage{newtxtext,newtxmath}
\usepackage{xcolor}
\usepackage{multirow}
\usepackage{mwe}
\usepackage{comment}
\usepackage{adjustbox}
\usepackage{graphicx}
\usepackage{amsmath}
\usepackage{threeparttable}
\usepackage{footnote}
\usepackage{tabularx}
\usepackage{graphicx}
\usepackage{soul}
\usepackage{epstopdf}
\usepackage{footnote}
\usepackage{float}
\usepackage{pdfpages}

\usepackage[utf8]{inputenc} % For UTF-8 encoding
\usepackage[T1]{fontenc} % For
\usepackage{amssymb}

\begin{document}

\begin{frontmatter}



\title{Exploring the magnetic field of the ultraluminous X-ray pulsar NGC 4631 X-8}


\author[1]{Amar Deo Chandra}
\ead{amar.deo.chandra@gmail.com}
\address[1]{Aryabhatta Research Institute of Observational Sciences, Manora Peak, Nainital, Uttarakhand, 263001, India}

\begin{abstract}
NGC 4631 X-8 is an ultraluminous X-ray pulsar (ULXP) having a spin period of about 9.7 s, discovered using \textit{XMM-Newton} observations in 2025. The pulsar is known to show one of the largest spin-up rates ($\sim -9.6 \times 10^{-8}$ \,s \,s$^{-1}$) among the ULXP population.
We explore the surface magnetic field of the neutron star in this source using different models, and find that the inferred magnetic field lies in the range of about $0.3-7 \times 10^{14}$G. We study the long-term magnetic field and spin period evolution of the pulsar assuming steady accretion using prevalent theoretical mechanisms and find that the pulsar will likely evolve to become a millisecond pulsar having decayed magnetic field of about  $\sim 10^{9}$G in about a million years.
The scenario of the formation of a millisecond pulsar is also probed using an estimate of the super-Eddington duty cycle of about 14\% from the literature, which suggests that the neutron star would accrete sufficient matter to likely become a recycled millisecond pulsar. Exploring the magnetic field as well as the spin period evolution properties of ULXPs may enable us to understand the poorly understood evolutionary features of ULXPs, shed light on one of the pathways of millisecond pulsar formation and also help us to understand the possibility of transient super-Eddington accretion phases in newborn magnetars, which are believed to power energetic events such as long gamma-ray bursts and Type I superluminous supernovae.
\end{abstract}


\end{frontmatter}

\section{Introduction}
\label{introduction}
The paradigm that Ultraluminous X-ray sources (ULXs) only hosted accreting stellar and intermediate-mass black holes \citep{kaaret2017ultraluminous} was challenged with the discovery of pulsations from some of these systems, suggesting that some ULXs may be powered by accreting neutron stars \citep{bachetti2014}. Some of the ultraluminous X-ray pulsars (ULXPs) are also transient Be/X-ray binary systems (\citet{chandra2020study} and references therein).
The elusive mechanism which powers super-Eddington luminosities in these systems is not well understood. One of the models suggests that the electron scattering cross-section is reduced due to strong magnetic fields (B$\sim 10^{14} ~G$), which may power super-Eddington luminosities in these objects \citep{canuto1971thomson,basko1976limiting,paczynski1992gb,dall2015nustar,ekcsi2015ultraluminous,mushtukov2015maximum,tong2015accreting}. Recent studies suggest that ULXPs may have magnetar-like strong dipolar magnetic fields (\citet{chandra2026} and references therein). However, it should be noted that although magnetars can have high magnetic fields (B$\sim 10^{14} ~G$), ULXPs and magnetars have different phenomenology and emission mechanisms. Another model suggests the presence of multipolar magnetic field components in these systems \citep{israel2017b,brice2021super}. The multipolar magnetic field model has been proposed for the most luminous ULXPs, as the magnetar-like field model suggests that ULXPs such as NGC 5907 ULX-1 would be in the propeller regime at the observed spin periods and luminosities.

The dipole magnetic field in ULXPs may be estimated using various methods. The accretion torque spinning up the neutron star, especially during outbursts \citep{chandra2020study,chandra2021detection,chandra2023astrosat}, is dependent on the interaction between the accretion disc and the dipole magnetic field \citep{ghosh1979accretion,kluzniak2007magnetically}. The critical luminosity (which marks the transition in the pulse profile of an X-ray pulsar from a fan beam to a pencil beam or vice-versa) is dependent on the dipole magnetic field strength of the pulsar \citep{becker2012spectral,mushtukov2015critical,chandra2023astrosat}. In addition, the torque balance equation and the minimum luminosity on the propeller line can also be used to estimate the dipole magnetic field \citep{christodoulou2016x,chandra2020study,chandra2025long,chandra2026}. The detection of cyclotron resonance scattering features (CRSFs) in the X-ray spectra of accreting pulsars provides the only direct estimation of the dipole magnetic field in these objects \citep{trumper1978evidence,wheaton1979absorption,kendziorra1994evidence,mihara1995observational,coburn2002magnetic,makishima2003measuring,staubert2019cyclotron,chandra2023astrosat}.

NGC 4631 X-8 is located in the starburst galaxy NGC 4631 at a distance of 7.6 Mpc \citep{monachesi2016ghosts}, which is known to host several ULXs \citep{soria2009different}. NGC 4631 X-8 is located about 7 arcseconds from the known ULX source NGC 4631 X-2 \citep{ducci2025discovery}. Pulsations of 9.6652$\pm$0.0002 s were discovered using \textit{XMM-Newton} observations in 2025 and the pulsar is known to exhibit one of the largest spin-up rates ($\sim -9.6 \times 10^{-8}$ \,s \,s$^{-1}$) among the confirmed ULXPs \citep{ducci2025discovery}.
The source was not detected using \textit{Chandra} observations from 2022 and 2023, and the estimated upper limit for the 0.2-12 keV luminosity was $\sim 4 \times 10^{36}$\,erg\,s$^{-1}$.
The inferred X-ray luminosity in the 0.3-10 keV energy band was $\sim 3.4 \times 10^{39}$\,erg\,s$^{-1}$ (d=7.6 Mpc) from \textit{XMM-Newton} observations in 2025 \citep{ducci2025discovery}, which confirmed that NGC 4631 X-8 is a new member of the class of transient ULXPs.

In this paper, we probe the surface dipole magnetic field of NGC 4631 X-8 using different models. The paper is organised as follows. After the introduction, we illustrate the different models in section 2 and estimate the dipolar surface magnetic field using these models. In section 3, we study the decay of the dipole magnetic field on long timescales. We explore the evolution of the magnetic field and spin period of the ULXP and compare its evolutionary pathway with known binary pulsars, magnetars and other ULXPs. Our findings are summarised in section 4.

\section{Magnetic field calculation}
Pulsations of 9.6652$\pm$0.0002 s were detected from NGC 4631 X-8 using \textit{XMM-Newton} observations and the inferred spin-up rate from timing analysis was $-9.6\pm 0.5\times 10^{-8}$ \,s \,s$^{-1}$ \citep{ducci2025discovery}, suggesting that the neutron star in this source accretes matter via an accretion disc as rapid spin-up during outbursts has been detected in other accreting pulsars wherein accretion is mediated through an accretion disc \citep{chandra2020study,chandra2021detection,chandra2023astrosat}. The accretion dynamics in accretion-powered pulsars is strongly regulated by the magnetospheric
(Alfv\'{e}n) radius ($r_{\rm m}$), which is given by \citep{davidson1973neutron},

\begin{equation}
r_{\rm m}=\xi\left(\frac{\mu^{4}}{2GM\dot{M}^{2}}\right)^{1/7}, \label{eq1}
\end{equation}

where $\xi$ is a dimensionless parameter typically assumed to be 0.5-1 \citep{monkkonen2019evidence}, $\mu=B_{\rm s}R^{3}$ is the surface magnetic dipole moment ($B_{\rm s}$ and R are the surface dipole magnetic field and the neutron star radius, respectively), $G$ is the gravitational constant, $M$ is the mass of the neutron star and $\dot{M}$ is the accretion rate at the Alfv\'{e}n radius. Using $R=10^6$ cm and $G=6.67\times 10^{-8}$~cm$^3$~g$^{-1}$~s$^{-2}$ we obtain,

\begin{equation}
r_{\rm m}=1.6\times 10^{8}\xi \dot{M}^{-2/7}_{18}M^{-1/7}_{1.4}\mu^{4/7}_{30}~\rm cm,
\end{equation}
where $\dot{M}_{18}$, $M_{1.4}$ and $\mu_{30}$ are in units of $10^{18}~\rm g\,s^{-1}$, $1.4~\rm M_{\odot}$, and $10^{30}~\rm G\,cm^{3}$ respectively.

The spin-up of neutron stars during outbursts can be ascribed
to the torque applied by the angular momentum imparted by the
accreted matter from the companion star and the magnetic coupling between the neutron star and the inner region of the accretion disc \citep{ghosh1979accretion}. The spin change in the neutron star due to magnetic torques is dependent on the fastness parameter
$\omega_{\rm s}=\Omega_{NS}/\Omega(r_{\rm m})$ where $\Omega_{NS}$, and $\Omega(r_{\rm m})$ are the angular velocity of the neutron star, and the Keplerian angular velocity at the magnetospheric radius, respectively. For $\omega_{\rm s} < 1$, the neutron star is spun-up by both the accreted matter and the magnetic interaction between the disc and the neutron star, while for $\omega_{\rm s} > 1$  the neutron star is spun down due to the magnetic interaction between the disc and the neutron star. Ignoring the contribution of the magnetic torque we obtain \citep{chen2017constraining,chandra2026},

\begin{equation}
-\frac{2\pi I\dot{P}}{P^{2}}\leq \dot{M}\sqrt{GMr_{\rm m}}, \label{eq2}
\end{equation}
where P, $\dot{P}$ and $I=\frac{2MR^2}{5}$ are the spin period, spin-up rate and the moment of inertia of the neutron star, respectively. Using $r_{\rm m}$ from equation \ref{eq1} in equation \ref{eq2} and $I \sim$ 1.9 $\times 10^{45}$\,g\,cm$^3$ (using $M=1.4 ~M_{\odot}$ and R$\sim$13 km), the lower limit on the dipole magnetic moment ($\mu_{\rm min,30}$) is obtained as,

\begin{equation}
\mu_{\rm min,30}\ge 8.96 \times 10^{37} \xi^{-7/4}\dot{M}^{-3}_{18}M^{-3/2}_{1.4} P^{-7} \dot{P}^{7/2}~\rm G\,cm^{3}.
\end{equation}

Using P$\sim$9.7 s and $\dot{P} \sim -9.6 \times 10^{-8}$ \,s \,s$^{-1}$ \citep{ducci2025discovery} for NGC 4631 X-8, the minimum dipole moment is given by,

\begin{equation}
\mu_{\rm min,30}=3.12 \times 10^{6} \xi^{-7/4}\dot{M}^{-3}_{18}M^{-3/2}_{1.4}~\rm G\,cm^{3}. \label{eq5}
\end{equation}

The maximum limit on the magnetic dipole moment can be obtained by equating the magnetospheric radius and the co-rotation radius of the neutron star ($r_{\rm co}$), which is given by,

\begin{equation}
r_{\rm co} =\biggl( {\frac{GMP^{2}}{4\pi^{2}}} \biggr)^{1/3}~\rm cm. \label{eq6}
\end{equation}

From equations \ref{eq1} and \ref{eq6} and assuming $r_{\rm m}=r_{\rm co}$, the maximum dipole magnetic moment ($\mu_{\rm max,30}$) is obtained as,

\begin{equation}
\mu_{\rm max,30}\le 1.09  \xi^{-7/4}\dot{M}^{1/2}_{18}M^{5/6}_{1.4} P^{7/6} ~\rm G\,cm^{3}. \label{eq7}
\end{equation}

Equating $\mu_{\rm max,30} \simeq \mu_{\rm min,30}$ and using M=$1.4~\rm M_{\odot}$, the minimum accretion rate at the magnetospheric radius is estimated as

\begin{equation}
\dot{M}_{\rm min}= 3.3\times 10^{19}~{\rm g\,s^{-1}}.
\end{equation}

The estimated minimum accretion rate $\dot{M}_{\rm min}$ is about 33 times the Eddington accretion rate $\dot{M}_{\rm Edd}$ ($\sim 1.0\times 10^{18}~\rm g\,s^{-1}$) for spherical accretion onto a neutron star. This implies that the source is likely to undergo super-Eddington accretion during outbursts.
In fact, the mass accretion rate estimated during the 2025 outburst of
this pulsar was $\dot{M} \sim 3\times 10^{19}~{\rm g\,s^{-1}}$ \citep{ducci2025discovery}, which
is comparable to that estimated earlier in this section.

\begin{figure}
\centering
  \includegraphics[width=\columnwidth]{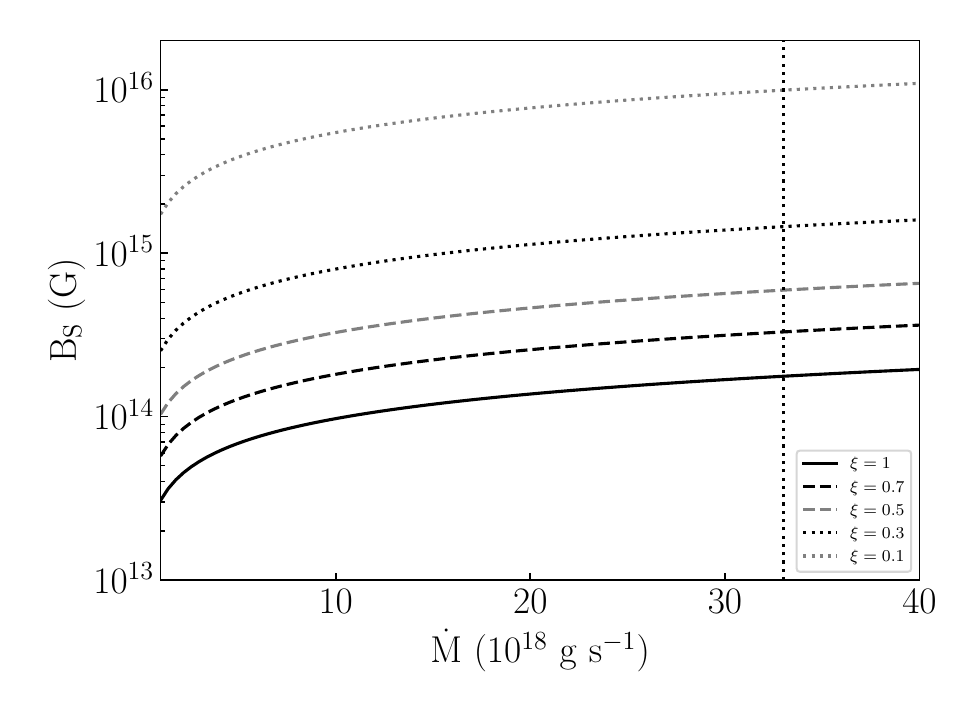}
  \caption{Variation of the surface dipole magnetic field of the neutron
star ($B_S$) and the accretion rate at the magnetospheric radius ($\dot{M}$).
The vertical dotted line marks the accretion rate of $3.3\times 10^{19}~{\rm g\,s^{-1}}$ estimated in section 2.}
 \label{f1}
\end{figure}

Using equation \ref{eq7}, $\mu=B_{\rm s}R^{3}$, $\xi \sim 1$, $\dot{M}_{18}=33$ and P$\sim$9.7 s the maximum surface dipole magnetic field of the neutron star ($B_{s,max}$) is estimated to be about $2 \times 10^{14}\,G$. Using $\xi =0.5$ (for disc accretion), the maximum surface dipole magnetic field of the neutron star ($B_{s,max}$) is estimated to be about $6.7 \times 10^{14}\,G$. The variation of the estimated surface dipole magnetic field of the neutron star with accretion rate at the magnetospheric radius is shown in Fig. \ref{f1}.

The magnetic field can also be estimated independently by using the torque balance equation \citep{christodoulou2016x,chandra2020study,chandra2025long,chandra2026},

\begin{equation}
B = \left(2\pi^2 \xi^7\right)^{-1/4}
\sqrt{\frac{G M I}{R^6} |\dot{P_S}|} \, ,
%\label{b1}
\end{equation}

where $\xi$ is a dimensionless parameter, which is the ratio of the inner edge of the accretion disc and the magnetospheric radius \citep{ghosh1979accretion,wang1996location,christodoulou2016x,chandra2020study, chandra2025long,chandra2026} and $\dot{P_S}$ is the rate of spin change of the pulsar during outbursts. Using $\xi$=1, $M=1.4 ~M_{\odot}$, R=10 km, $\dot{P_S} \sim -9.6 \times 10^{-8}$ \,s \,s$^{-1}$ \citep{ducci2025discovery} and $G=6.67\times 10^{-8}$~cm$^3$~g$^{-1}$~s$^{-2}$, the estimated magnetic field of the neutron star is $\sim 6.7 \times 10^{13} ~G$. Using $\xi =0.5$ (for disc accretion), the estimated magnetic field of the neutron star is $\sim 2.3 \times 10^{14} ~G$.
%, which is a factor of about 3 less than that estimated earlier.
The magnetic field can also be estimated independently from the minimum luminosity on the propeller line \citep{christodoulou2016x,chandra2020study,chandra2025long,chandra2026},

\begin{equation}
B = 8.0\times 10^{11} \sqrt{\frac{L_{X,38}}{\eta}}\left(\frac{P_S}{1~{\rm s}}\right)^{7/6} ~~{\rm G}\, ,
\label{pline4}
\end{equation}

where $L_{X,38} = L_X/1.0\times 10^{38}~{\rm erg~s}^{-1}$. Using $P_S \sim 9.7$~s, $L_{X,38} \sim 1.8$, and $\eta = 1/4$ \citep{christodoulou2016x,chandra2020study,chandra2025long,chandra2026}, the estimated magnetic field is $\sim 3 \times 10^{13} ~G$.

The magnetic field of the neutron star estimated using the model given by \cite{ghosh1979accretion} is $\sim 10^{13} ~G$ \citep{ducci2025discovery}. \citet{ducci2025discovery} also estimate the magnetic field using the model given by \citet{gao2021magnetic}, which extends the magnetically threaded disk model to the super-Eddington regime. Using this model, the estimated magnetic fields for regimes far away and close to spin equilibrium are $\sim 10^{13} ~G$ and $\sim 2 \times 10^{14}\,G$, respectively \citep{ducci2025discovery}. The higher magnetic field inferred by \citet{ducci2025discovery} is equal and about threefold smaller than the magnetic field obtained using equation \ref{eq7} for $\xi \sim 1$ and $\xi =0.5$, respectively. The lower value of the magnetic field ($\sim 10^{13} ~G$), is comparable (within a factor of about 3-23) to those estimated using the torque balance equation and the minimum luminosity on the propeller line.

\section{Magnetic field evolution}
The magnetic field evolution of an accretion-powered neutron star has been explored in several studies \citep{bisnovatyi1974pulsars,taam1986magnetic,shibazaki1989does,geppert1994accretion,van1995magnetic,melatos2000evolution,zhang2006bottom,heuvel2009formation,ho2011evolution,igoshev2021evolution,pan2016magnetic,pan2022study,meng2022research,chandra2026}. We explore the magnetic field evolution of NGC 4631 X-8 using the model of accretion-induced magnetic field decay of a neutron star (\citet{zhang2006bottom}, referred to as the ZK model). This model has been used to study the magnetic evolution of several accreting pulsars such as NuSTAR J095551+6940.8, NGC 300 ULX1, NGC 7793 P13, and RX J0209.6-7427 \citep{pan2016magnetic,pan2022study,meng2022research,chandra2026}.
The evolved magnetic field in the ZK model is given by the analytic solution \citep{zhang2006bottom},

\begin{equation}
\label{eq15}
B=\frac{B_{\rm f}}{\{1-[C/{\rm exp}(y)-1]^2\}^{7/4}},
\end{equation}
where $B_{\rm f}$ is the bottom magnetic field ($B_{\rm f}\simeq 1.32\times10^8(\dot M/\dot M_{\rm Edd})^{1/2}m^{1/4}R_6^{-5/4}\phi^{-7/4}\,\rm G$, ), $m=1.4~\rm M_{\odot}$  is the mass of the neutron star, R=10 km is the typical neutron star radius, and $\phi$ is the ratio of the magnetospheric radius to the Alfv$\acute{\rm e}$n radius, which is assumed to be 0.5. The bottom magnetic field refers to the magnetic field at the end of the evolution.
$C=1+(1-X_0^{2})^{1/2}\sim 2$ where $X_0^{2}=(B_{\rm f}/B_0)^{4/7}$, $B_0$ is the initial magnetic field of the neutron star. The parameter $y=2\mathbf{\epsilon}\Delta M/7M_{\rm cr}$ is the ratio of the accreted mass ($\Delta M$) to the crust mass ($M_{\rm cr}\sim0.2\,M_{\odot}$) of a neutron star. The mass accreted by a neutron star is given by $\Delta M=\dot M T_{\rm ac}$, where $T_{\rm ac}$ is the accretion time. The efﬁciency factor $\mathbf{\epsilon}$ is taken as unity for completely frozen ﬂow of the magnetic line due to the plasma instability.

\begin{figure}
\centering
  \includegraphics[width=\columnwidth]{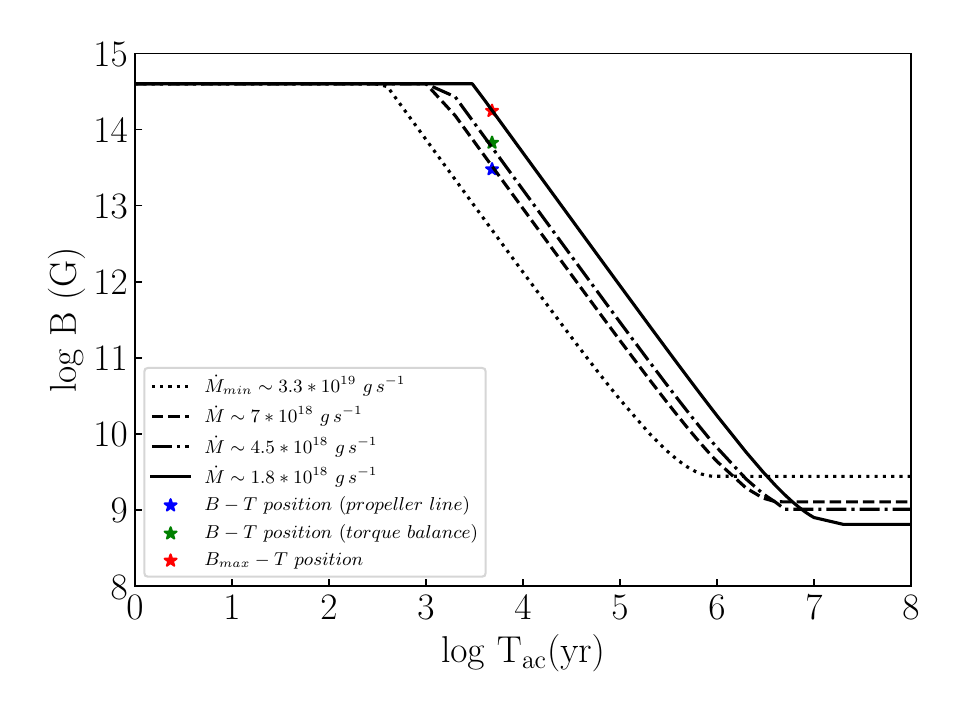}
  \caption{Accretion-induced decay of the magnetic field of NGC 4631 X-8. The magnetic field evolution is assumed to begin at $T_{ac}=0$ with initial  $B_{0}=4 \times 10^{14}\,G$. The curves plotted with dotted, dashed, dashed-dotted, and solid lines show the magnetic field evolution for $\dot{M_{min}}\sim 3.3\times 10^{19}~{\rm g\,s^{-1}}$, $\dot{M}\sim 7\times 10^{18}~{\rm g\,s^{-1}}$, $\dot{M}\sim 4.5\times 10^{18}~{\rm g\,s^{-1}}$, and $\dot{M}\sim 1\times 10^{18}~{\rm g\,s^{-1}}$, respectively. The magnetic field decays to the bottom value of $\sim 0.7-1.5 \times 10^{9}\,G$ in about $10^{6-7}$ years. The $B-T_{ac}$ positions are plotted for the minimum and the maximum estimated values of the magnetic field.}
 \label{f2}
\end{figure}

The evolution of the magnetic field is simulated for $\dot{M}_{min}\sim 3.3\times 10^{19}~{\rm g\,s^{-1}}$, $\dot{M}\sim 1.8\times 10^{18}~{\rm g\,s^{-1}}$, $\dot{M}\sim 4.5\times 10^{18}~{\rm g\,s^{-1}}$, and $\dot{M}\sim 7\times 10^{18}~{\rm g\,s^{-1}}$ as shown in Fig. \ref{f2}. It is unlikely that the pulsar will always accrete at a high accretion rate of $\dot{M}\sim 3.3\times 10^{19}~{\rm g\,s^{-1}}$, and so we have chosen a few other trial values of accretion rate in the super-Eddington regime to explore the magnetic evolution of the source. It is assumed that the evolution of the dipolar magnetic field begins at $T_{ac}=0$ and $B_{0}=4 \times 10^{14}\,G$. The initial magnetic field decays to the present estimated magnetic field of about $0.3-2 \times 10^{14}\,G$ in about $5 \times 10^3$ years. The magnetic field will decay to the bottom magnetic ﬁeld of $\sim 0.7-2.8 \times 10^{9}\,G$ in about $10^{6-7}$ years after accreting mass of about $0.3-0.5~\rm M_{\odot}$. The accretion time ($T_{ac}$) of a neutron star in a binary system is related to the lifetime of the companion star in the main sequence, given by \citep{shapiro1983black},

\begin{equation}
T_{\rm ac}=1.3\times10^{10}\, f \,m_{\rm c}^{-2.5}\,\rm yr,
\label{eq16}
\end{equation}
%%%%%%
where $m_{\rm c}$ is the mass of the companion star (in units of solar mass) and $f$ is the accretion efficiency factor, which is about 0.1 \citep{shapiro1983black}.
The estimated mass of the companion star in NGC 4631 X-8 is about 15-70 $\rm M_{\odot}$ \citep{ducci2025discovery}. The corresponding accretion time is estimated to be about $0.03-1.5\times 10^{6}$ years using equation \ref{eq16}. The magnetic field of NGC 4631 X-8 will decay to $\sim 0.001-1.5 \times 10^{13}\,G$ within this accretion time (assuming $\dot{M}\sim 1.8\times 10^{18}~{\rm g\,s^{-1}}$) limited by the mass of the companion star. The accreted mass from the companion star during this duration will be about 0.001-0.04 $\rm M_{\odot}$ for an accretion rate of $\sim 1.8\times 10^{18}~{\rm g\,s^{-1}}$. Assuming an accretion rate of $\sim 3.3\times 10^{19}~{\rm g\,s^{-1}}$, the magnetic field will decay to $\sim 0.03-2 \times 10^{11}\,G$ and the accreted mass from the companion star will be about 0.02-0.78 $\rm M_{\odot}$ within this accretion time ($0.03-1.5\times 10^{6}$ years).

\begin{figure}
\centering
  \includegraphics[width=\columnwidth]{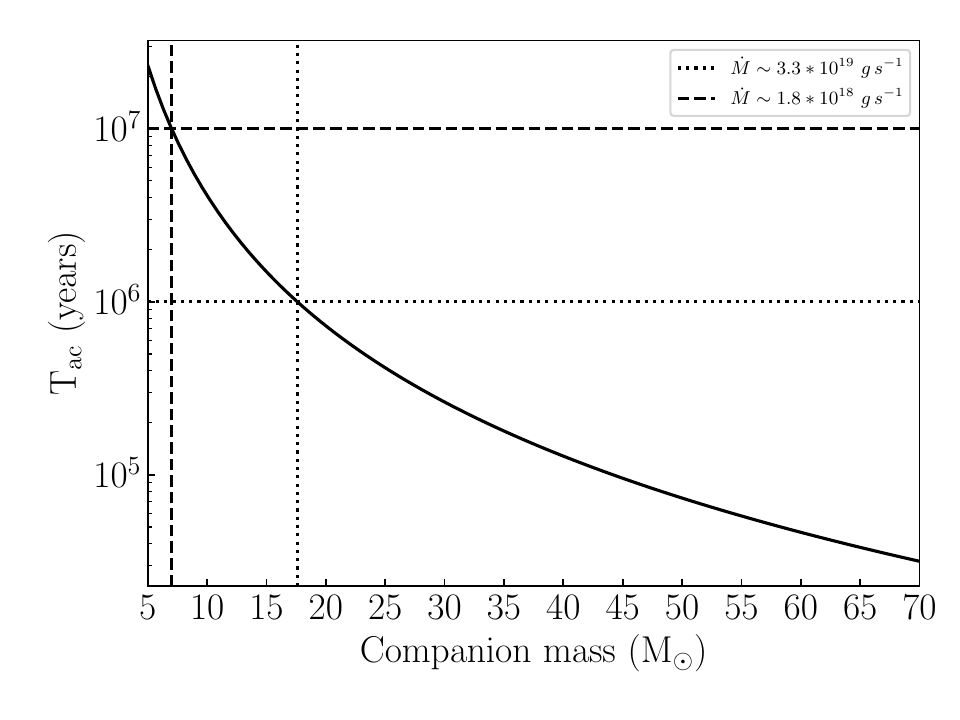}
  \caption{Accretion time ($T_{\rm ac}$) variation with mass of the companion star. The dotted and dashed lines show the estimated accretion time and the companion mass for $\dot{M}\sim 3.3\times 10^{19}~{\rm g\,s^{-1}}$, and $\dot{M}\sim 1.8\times 10^{18}~{\rm g\,s^{-1}}$, respectively.}
 \label{f3}
\end{figure}

The variation of the accretion time ($T_{\rm ac}$) with mass of the companion star (using equation \ref{eq16}) is shown in Fig. \ref{f3}. The estimated accretion time for a steady accretion rate of $\dot{M}\sim 3.3\times 10^{19}~{\rm g\,s^{-1}}$, and $\dot{M}\sim 1.8\times 10^{18}~{\rm g\,s^{-1}}$ is $\sim 10^{6}$ years and $\sim 10^{7}$ years, respectively as shown in Fig. \ref{f3}. The companion masses estimated for accretion time of about $\sim 10^{6}$ years and $\sim 10^{7}$ years is about 18 $\rm M_{\odot}$ and 7 $\rm M_{\odot}$, respectively. The mass of the companion star in other confirmed ULXPs vary in the range of about 5-23 $\rm M_{\odot}$ (Table \ref{t3}), which suggests that the mass of the companion in NGC 4631 X-8 may be about a few tens of solar mass, which needs to be confirmed from optical observations.

\begin{table*}
\caption{\normalsize{Parameters of known ultraluminous X-ray pulsars \small(\url{https://amarastro.github.io/ULX_pulsars.html})}} % title of Table
\label{t3}
\centering % used for centering table
\begin{tabular}{c c c c c c c c} % centered columns (4 columns)
\hline\hline %inserts double horizontal lines
\normalsize{Source} & \normalsize{d (Mpc)} & \normalsize{Spin period (s)} & \normalsize{$P_{orb}$(d)} & \normalsize{$L_X ~(10^{39}$\,ergs\,s$^{-1})$} &  \normalsize{Donor mass ($\rm M_{\odot}$)} & \normalsize{$B_{13}$ (G)} & \normalsize{References}\\  % inserts table
%       &     &            &     &     &\\[0.5ex]

%heading
\hline % inserts single horizontal line

 \normalsize{M82 X-2} &  \normalsize{3.6} & \normalsize{1.37} & \normalsize{$\sim 2.5$} & \normalsize{$\sim$20} &  \normalsize{$\gtrsim$ 5.2} & \normalsize{$\sim$2-3} & \normalsize{1,2,3}\\
 \normalsize{NGC 7793 P13} &  \normalsize{3.9} & \normalsize{$\sim 0.42$} & \normalsize{64} & \normalsize{$\sim 10$} &  \normalsize{$\sim$18-23} & \normalsize{$\sim$0.2-0.5} & \normalsize{3,4,5,6}\\
 \normalsize{NGC 5907 ULX} &  \normalsize{17.1} & \normalsize{$\sim 1.13$} & \normalsize{5.3} & \normalsize{$\sim 100$} &  \normalsize{$\gtrsim$ 20} & \normalsize{$\sim$0.2-3}& \normalsize{7}  \\
 \normalsize{NGC 300 ULX1} &  \normalsize{1.88} & \normalsize{$\sim 31.6$} & \normalsize{-} & \normalsize{4.7} &  \normalsize{$\sim$8-10} & \normalsize{$\sim$30}& \normalsize{8,9,10}\\
\normalsize{M51 ULX-7} &  \normalsize{8.6} & \normalsize{$\sim 2.8$} & \normalsize{$\sim 2$} & \normalsize{$\sim 10$} &  \normalsize{$\gtrsim$8} & \normalsize{$\sim$2-7}&  \normalsize{11,12}\\
 \normalsize{NGC 1313 X-2} &  \normalsize{4.2} & \normalsize{$\sim 1.5$} & \normalsize{-} & \normalsize{$\sim 20$} &  \normalsize{$\lesssim$12} & \normalsize{-}&  \normalsize{13}\\

 \normalsize{\textit{Swift} J0243.6+6124} &  \normalsize{0.007} & \normalsize{$\sim 9.86$} & \normalsize{$\sim 27.6$} & \normalsize{$\sim 2$} &  \normalsize{16} & \normalsize{$\sim$1.6}& \normalsize{14,15,16}\\
 \normalsize{RX J0209.6-7427} &  \normalsize{0.06} & \normalsize{$\sim 9.3$} & \normalsize{-} & \normalsize{$\sim 1.6$} &  \normalsize{-} & \normalsize{$\sim$2.4-4}&  \normalsize{17,18}\\
   \normalsize{SMC X-3} &  \normalsize{0.06} & \normalsize{$\sim 7.78$} & \normalsize{$\sim 45$} & \normalsize{$\sim 2.5$} &  \normalsize{$\gtrsim$ 3.7} & \normalsize{$\sim$2-3}& \normalsize{19}\\
 \normalsize{NGC 4631 X-8} &  \normalsize{7.5} & \normalsize{$\sim 9.67$} & \normalsize{-} & \normalsize{$\sim 3.4$} &  \normalsize{$\sim$15-70} & \normalsize{$\sim$3-70}& \normalsize{20,21}\\
\hline %inserts single line
\end{tabular}
\label{table:nonlin} % is used to refer this table in the text
\\\normalsize{(1) \citet{bachetti2014},} (2) \citet{dall2015nustar}, (3) \citet{gao2021magnetic}, (4) \citet{furst2016}, (5) \citet{israel2017a}, (6) \citet{motch2014mass}, (7) \citet{israel2017b}, (8) \citet{carpano2018}, (9) \citet{heida2019discovery},  (10) \citet{pan2022study}, (11) \citet{castillo2019}, (12) \citet{vasilopoulos2020m51}, (13) \citet{sathyaprakash2019}, (14) \citet{wilson2018}, (15) \citet{reig2020optical}, (16) \citet{kong2022insight}, (17) \citet{chandra2020study}, (18) \citet{chandra2026}, (19) \citet{tsygankov2017smc},  (20) \citet{ducci2025discovery}, \normalsize{and (21) this work.}
\end{table*}

\begin{figure}
\centering
  \includegraphics[width=\columnwidth]{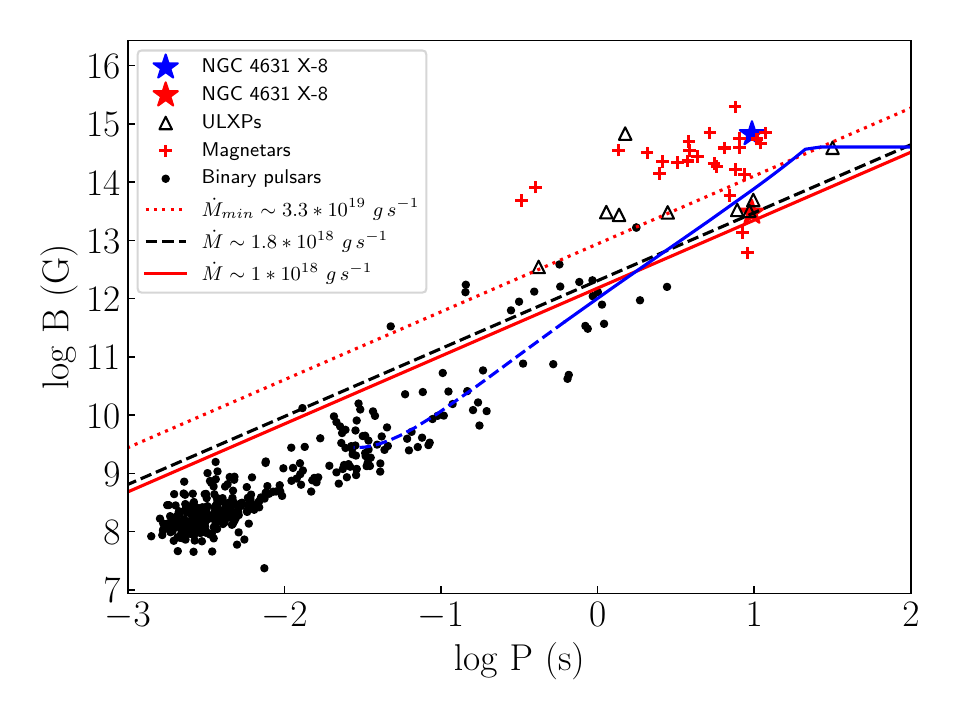}
  \caption{Diagram showing magnetic field vs spin period. The red and blue stars in the figure show the B-P position of NGC 4631 X-8 for $B_{min} \sim 3 \times 10^{13}\,G$ and $B_{max} \sim 6.7 \times 10^{14}\,G$, respectively, having spin period of about 9.67 s. The dotted, dashed and solid lines exhibit the spin-up line having accretion rate $\dot{M}\sim 3.3\times 10^{19}~{\rm g\,s^{-1}}$, $\dot{M}\sim 1.8\times 10^{18}~{\rm g\,s^{-1}}$, and the Eddington accretion rate, respectively. The evolved B-P path is shown by the blue line. The B-P positions of binary pulsars obtained from the ATNF pulsar catalogue (\url{https://www.atnf.csiro.au/research/pulsar/psrcat/}, \citet{manchester2005australia}) are marked by solid dots while those of magnetars are marked by plus signs. The B-P positions of other confirmed ultraluminous X-ray pulsars M82 X-2, NGC 7793 P13, NGC 5907 ULX, NGC 300 ULX-1, M51 ULX-7, NGC 1313 X-2, Swift J0243.6+6124, RX J0209.6-7427, and SMC X-3 obtained from literature are plotted with triangles.}
 \label{f4}
\end{figure}

The evolution of the dipolar magnetic ﬁeld and the spin period of NGC 4631 X-8 is simulated using the following equation \citep{ye2020exploring}

\begin{equation}
\frac{2\pi}{P}=\frac{2\pi}{P_0}+ \mathbf{\alpha} {B^{2/7}_{012}} \Bigg[\Bigg(1+\frac{\dot{M} t}{m_B}\Bigg)^{5/7}-1\Bigg],\label{eq13}
\end{equation}

where $P_0$ is the initial spin period of the neutron star, $\mathbf{\alpha}=\frac{7}{5}k \Big(\frac{m_B}{\dot{M}}\Big)$ where k is given by $k=1.87\times10^{-10}\dot M_{18}^{6/7}m^{-4/7}R_6^{-8/7}\phi^{1/2}$, $B_{012}$ is the initial dipole magnetic field of the neutron star and $m_B=0.5(B_{\rm f}/B_0)^{4/7}M_{cr}$ \citep{zhang2006bottom}. We assume steady accretion with the defined accretion time and rate. The accretion time is assumed to be the Hubble time $1.3 \times 10^{10}$ yr \citep{ade2016planck}. The accretion rate used is $\dot{M}\sim 3.3\times 10^{19}~{\rm g\,s^{-1}}$. We use $P_0=100$ s, $B_{012}$=400 G, m=1.4 $\rm M_{\odot}$, $R=10^6$ cm, $\phi$=0.5, $M_{\rm cr}=0.2\,M_{\odot}$ and $B_f=2.8 \times 10^{9}\,G$. The evolved B-P path (computed using equations \ref{eq15} and \ref{eq13}) is shown in Fig. \ref{f4} by the blue line, where the dashed part (at the end of the track) is
limited by the accretion time of $0.03-1.5\times 10^{6}$ yr calculated earlier. The B-P evolution of NGC 4631 X-8 shows a plateau at the beginning, as the magnetic field does not decay during this period and remains fixed at the initial magnetic field ($B_{012}$=400 G). The plateau-like feature seen near the end of the B-P evolution is due to the decayed magnetic field approaching the bottom magnetic field. The spin-up or the equilibrium period lines shown in Fig. \ref{f4} for accretion rate of $\sim 3.3\times 10^{19}~{\rm g\,s^{-1}}$, $\sim 1.8\times 10^{18}~{\rm g\,s^{-1}}$,  and the Eddington accretion rate are plotted using the minimum spin period corresponding to a given magnetic field assuming that the spin frequency is the same as the Keplerian frequency at the magnetospheric radius \citep{bhattacharya1991formation}. It is observed that NGC 4631 X-8 will likely evolve to become a millisecond pulsar having a spin period of about hundred milliseconds at the end of the accretion phase.

\section{Discussions}
The magnetic field of other confirmed ultraluminous X-ray pulsars has been estimated in several studies using different models \citep{erkut2020magnetic,chen2021studying,gao2021magnetic,meng2022research,chandra2026}.
It has been suggested that
ultraluminous X-ray pulsars are likely to have strong dipolar surface magnetic fields (Table \ref{t3}). Among the ULXP population, cyclotron resonance scattering features during outbursts have been detected only in one source viz.  \textit{Swift} J0243.6+6124. A CRSF around 146 keV was detected in this source, which was used to estimate the surface magnetic field of $\sim 1.6 \times 10^{13}$G \citep{kong2022insight}.
Thus, detections of CRSFs in other ULXPs is needed to directly infer the magnetic field of the neutron star in these sources and corroborate the paradigm that ULXPs have strong dipolar magnetic fields ($B\sim 10^{13}\,G$). M51 ULX-8 is a candidate ULXP in which an absorption line at 4.5 keV has been detected in the \textit{Chandra} spectrum \citep{brightman2018magnetic}, and it is still a second (possible) detection of CRSF in an ULX(P).

Twelve binary pulsars listed in the ATNF pulsar catalogue  \citep{manchester2005australia} lie above the Eddington spin-up line in the B-P diagram (Fig. \ref{f4}). It is likely that these sources may have experienced super-Eddington accretion during their evolutionary phase. Studies suggest that super-Eddington accretion at some point in the evolutionary phase of neutron star binaries may spin up the neutron stars in these systems to millisecond periods \citep{webbink1997super,lazarus2014timing,tauris2017formation,deng2025formation,guo2025eclipsing}. The B-P positions of confirmed ultraluminous X-ray pulsars (Fig. \ref{f4}), suggests that most of the ULXPs are accreting pulsars having spin periods in the range of about 1-30 s and likely having strong dipolar magnetic fields ($B\gtrsim 1 \times 10^{13}\,G$). Most of the magnetars lie above the Eddington spin-up line in the B-P diagram (Fig. \ref{f4}), with two magnetars just below the Eddington spin-up line. However, this is not necessarily a sign of previous super-Eddington accretion in magnetars, since these objects are powered by a different mechanism than the other sources shown in Fig. \ref{f4}. Recent studies suggest that newborn magnetars undergoing super-Eddington accretion may be one of the possible central engines of some long gamma-ray bursts, Type I superluminous supernovae or even failed supernovae \citep{stratta2018magnetar,zhong2019gravitational,lin2020unified,xie2022constraining,lan2026constraining}. It has also been suggested that newborn magnetars accreting at super-Eddington rates may be potential sources of gravitational wave emission \citep{piro2012gravitational,zhong2019gravitational,sur2021gravitational}. However, it should be noted that super-Eddington accretion in the early stages of a magnetar's life is still debated.

The evolutionary track of NGC 4631 X-8 in the B-P diagram crosses the zone of
magnetars and binary pulsars above the Eddington spin-up line and then traverses through the zone of recycled binary pulsars and finally ends around the positions of recycled millisecond pulsars at the end of the accretion phase of the binary. The estimated spin period and the magnetic field of NGC 4631 X-8 (assuming steady accretion of $\dot{M}\sim 3.3\times 10^{19}~{\rm g\,s^{-1}}$) at the end of the accretion phase is about 100 milliseconds and  $\sim 3 \times 10^{9}\,G$, respectively.  It has been shown that accelerating a neutron star's spin period to about 100 ms requires the total accreted matter of at least 0.001$\rm M_{\odot}$ \citep{tauris2012formation}. This is within a factor of about 1-800 compared to the range of total accreted mass by the neutron star of about 0.001-0.04 $\rm M_{\odot}$ and 0.02-0.78 $\rm M_{\odot}$ for accretion rate of $\sim 1.8\times 10^{18}~{\rm g\,s^{-1}}$ and $\sim 3.3\times 10^{19}~{\rm g\,s^{-1}}$, respectively during the accretion phase of the binary ($0.03-1.5\times 10^{6}$ years), which is limited by the mass of the donor star (15-70 $\rm M_{\odot}$). It should be noted that a steady accretion rate has been assumed for calculating the total accreted mass by the neutron star, which may not be the case as the source is known to show long-term variability by three orders of magnitude \citep{ducci2025discovery}.  In a recent work, \citet{chandra2026} explored the B-P evolution of another ULXP RX J0209.6-7427 and found that the source would likely evolve to become a millisecond pulsar having a spin period and magnetic field of about a hundred milliseconds and $\sim 10^{9}\,G$, respectively, assuming a steady accretion rate of $\sim 6.6\times 10^{18}~{\rm g\,s^{-1}}$. The total matter accreted by the neutron star during the evolutionary phase of the binary lasting about
$0.5-3\times 10^{6}$ years (limited by the mass of the donor star of about 11-23 $\rm M_{\odot}$) was estimated to be about 0.05-0.3 $\rm M_{\odot}$
\citep{chandra2026}. The estimated accreted mass for accelerating a neutron star's spin period to about 50 ms is at least 0.001$\rm M_{\odot}$\citep{tauris2012formation}, which is about a factor of 50-300, smaller than that estimated for RX J0209.6-7427. RX J0209.6-7427 is also a transient ULXP with only two detections since its discovery in 1993 \citep{chandra2020study}. This suggests that the total accreted mass by the neutron star during the accretion phase of the binary in NGC 4631 X-8 and RX J0209.6-7427 is overestimated, as the duty cycle of outbursts in ULXPs is poorly understood.

In a theoretical study, \citet{tong2019accreting} estimated the upper limit on the duty cycle of ULXPs to be 14\%. Assuming a duty cycle of 14\%, the range of total accreted mass by the neutron star during the accretion phase in NGC 4631 X-8 is about 0.0001-0.006 $\rm M_{\odot}$ and 0.003-0.11 $\rm M_{\odot}$ for accretion rate of $\sim 1.8\times 10^{18}~{\rm g\,s^{-1}}$ and $\sim 3.3\times 10^{19}~{\rm g\,s^{-1}}$, respectively. These estimated mass ranges are within a factor of about 0.1-100 of the minimum required accreted matter by the neutron star of
0.001 $\rm M_{\odot}$ to accelerate the neutron star to a spin period of about 100 ms \citep{tauris2012formation}. In case of RX J0209.6-7427, assuming a duty cycle of 14\%, the range of total accreted mass by the neutron star during the accretion phase is about 0.007-0.04 $\rm M_{\odot}$for accretion rate of $\sim 6.6\times 10^{18}~{\rm g\,s^{-1}}$. This estimated mass range is within a factor of about 7-40 of the minimum required accreted matter by the neutron star of 0.001$\rm M_{\odot}$ to recycle the neutron star to a spin period of about 50 ms \citep{tauris2012formation}. Detection of the spectral type of the companion star and tighter constraints on its mass can enable us to constrain the duration of the accretion phase of the binary and provide better estimates of the total accreted matter by the neutron star during the accretion phase of the binary. Studies of the evolution of ULXPs may be helpful to understand magnetar evolution, the possibility of transient super-Eddington accretion phases in newborn magnetars, which are believed to power energetic events such as long gamma-ray bursts and Type I superluminous supernovae, evolution of accreting millisecond pulsars harbouring strong magnetic fields, and formation of binary pulsars and millisecond pulsars.

\section{Conclusions}
We probe the surface magnetic field strength of NGC 4631 X-8 employing different accretion models and estimate magnetic field of $\sim 0.3-7 \times 10^{14}$G. The magnetic field decay of the neutron star in the binary is simulated assuming steady accretion rates, and it is observed that the magnetic field will likely decay to $\sim 10^{9}$G at the end of the accretion phase of the binary in about a million years. The magnetic field and spin period evolution of the binary are simulated using constraints such as a steady accretion rate and accretion time dependent on the mass of the companion star. It is observed that the neutron star in NGC 4631 X-8 can accrete sufficient mass during the accretion phase of the binary to likely become a recycled millisecond pulsar with a spin period of about hundred milliseconds and that its magnetic field would likely decay to about $\sim 10^{9}$G, respectively. This scenario of the formation of a millisecond pulsar is also explored using the estimate of duty cycles of the ultraluminous phase in ULXPs. Future multiwavelength observations of ULXPs can be helpful to constrain the duty cycle, mass of the companion star, and hence the duration of the accretion phase in these binary systems, which can enable us to better constrain the spin and magnetic evolution of these sources.

\section*{Declaration of competing interest}
The authors declare that they have no known competing financial
interests or personal relationships that could have appeared to influence
the work reported in this paper.

\section*{Acknowledgements}
We thank the reviewer for their constructive suggestions that helped
to improve the manuscript.
This work is dedicated to Prof. Jayant Vishnu Narlikar (1938-2025), who first
inspired many of us as children to learn about astronomy through his iconic TV series
\textquotedblleft Brahmand-The Universe\textquotedblright.
This research has made use of NASA's Astrophysics Data System. ADC acknowledges support from ARIES through post-doctoral fellowship.

\section*{Data availability}
No new data were generated or analysed in support of this research.

\bibliographystyle{elsarticle-harv}
\bibliography{main}

\end{document}